# Electronic structures of impurities and point defects in semiconductors

Yong Zhang*

[1]Department of Electrical and Computer Engineering,

University of North Carolina at Charlotte, Charlotte, NC 28223, USA

## Abstract

A brief history of the impurity theories in semiconductors is provided. A bound exciton model is proposed for both donor- and acceptor- like impurities and point defects, which offers a unified understanding for "shallow" and "deep" impurities and point defects. The underlying physics of computational results using different density-functional theory based approaches are discussed and interpreted in the framework of the bound exciton model.

* yong.zhang@uncc.edu

# 1. Introduction

Impurities and point defects are very similar in terms of their primary functions in semiconductors. For instance, they both can behavior as electron donors or acceptors to change the electrical conductivity of the material. In terms of theoretical treatments, the impurity and defect problem are also very similar, namely, one host atom on one particular lattice site is replaced by another atom that normally should not be there in the perfect lattice. A few typical examples are offered here to illustrate the point that most impurities or point defects behavior either as a donor or an acceptor in a semiconductor. (1) In Si, Al substituting Si results in an acceptor state, because Al has one less valence electron than Si and Al 3p state is higher than Si 3p state, an unoccupied impurity state is likely to appear above the top of the valence band or valence band maximum (VBM). (2) In GaAs, a Ga on As anti-site defect is expected to behavior as an acceptor, because the As site replaced by Ga is short of two valence electrons and Ga 4p state is higher than As 4p state, an unoccupied defect state is likely to appear above the VBM. Here the Ga on the wrong lattice site can be viewed as either an anti-site defect or an impurity. (3) In GaP, an N impurity substituting P is another example of an acceptor impurity, which is often known as an isoelectronic impurity, because N and P have the same number of valence electrons. Because N 2s state is lower than P 3s state, an N impurity has the tendency to form an empty level (a s-like anti-bonding state) below the bottom of the conduction band or conduction band minimum (CBM), which can be viewed as a deep acceptor. By deep acceptor, we mean that the acceptor level is very far away from the VBM for the electron in the valence band to make a transition to the acceptor level. (4) In NaCl, a Cl vacancy behaves like a deep donor, because in this ionic crystal, Na atoms are supposed to give out their valence electrons to Cl atoms; now with one Cl missing, one extra valence electron of the nearby Na atoms has to find a state to occupy. This state turns out to be localized at the Cl vacancy. This defect state is a deep donor state, i.e., the bound electron has a large binding energy with respect to the CBM.

The Cl vacancy, known as a "color center" or "F center", played a very important role in the history of the impurity and defect theory in semiconductors. The well-known hydrogen model was first proposed by Gurney and Mott [1] and Tibbs [2] to understand the electronic structure of this simple point defect. For the extra valance electron, if somehow the vacancy site can still keep this electron, the general volume of the crystal, away from the vacancy site, will more or less remains the same as the defect-free crystal. This arrangement is indeed possible, and it is normally considered as a neutral state of the vacancy ($V^0$). However, if the electron is released from the vacancy, for instance, being excited into the conduction band, the crystal will exhibit some conductivity, and we may say that the Cl vacancy is ionized. In the latter case, relative to the charge distribution of the defect-free crystal, the defect site has a positive charge, thus, the vacancy is said to be in +1 charge state ($V^+$). One could view a vacancy as a virtual atom that has an empty electronic state at the vacuum level, which suggests that the vacancy site tends to push away the electron, acting like an anti-quantum-dot. Alternatively, a vacancy could also be viewed as an interface between the vacuum and crystal "surface" with dangling bonds. The dangling bonds in a semiconductor surface is known to often generate a surface state that is highly localized at the surface. Despite the vacuum space is quite small, a highly localized state can indeed form in the small cavity to accommodate the "orphan" electron. Because this state is an anti-bonding state in nature (to be explained later), its energy level is mostly likely to be close to the conduction band, which is a rather general phenomenon for an anion vacancy in an ionic crystal [3]. The examples given above illustrate that at least most impurity and defect problems can be understood qualitatively in a similar way by considering the electronic structure difference between the host and "impurity" atom, with the help of the knowledge about the host band structure, to predict if the "impurity" should behave as an acceptor or a donor. This point is particularly important for the introduction of a new and unified theoretical framework [4] for the impurity and point defect problem in this paper. Therefore, in the discussions below, the word "impurity" may be understood as representing either impurity or point defect, unless specifically stated otherwise.

In the literature, impurities are typically classified into two categories: "shallow" and "deep". In the early days, an impurity is deemed as "shallow" when the separation of its ground state energy level from the relevant band edge, i.e., "impurity binding energy", is comparable to the thermal energy kT corresponding to room temperature (the usual device operating temperature), and as "deep" otherwise [5]. This intuitive classification is of practically useful, because from the device operation point of view, the exchange of electrons between the impurity levels and the bulk band states depends sensitively on the "impurity binding energy". The exact meaning of the term "impurity binding energy" will be the subject of later discussions. However, another criterion of classification of "shallow" and "deep" impurities have also been widely used. It emphasizes the difference in the degree of impurity potential localization by recognizing the fact that an impurity level despite being generated by a highly localized impurity potential can be energetically very close to the band edge, but its properties can be very different from the specific band edge [6]. The best example may be the electron bound state of an isolated N impurity in GaP, GaP:N, that has an impurity level very close to the lowest conduction band state near the X point, but the pressure response of the exciton bound to the N center is found to be very different from that of the X point. Another well cited example is a resonant state of an isolated N impurity in GaAs, GaAs:N, that has an impurity level well above the conduction band edge at the Γ point and somewhat close to the conduction band L point, but the pressure response of this impurity level does not follow any of three critical points at Γ, L, and X. The qualitative explanations for these two examples are relatively simple: an impurity state associated with a highly localized potential will require states throughout the BZ, maybe even from different bands, to serve as a basis for its wave function expansion using the host band states, thus, one cannot expect the impurity state to behave like one particular band edge state, even they could be incidentally close to each other. It was because of the examples like these two isoelectronic impurity systems, it was proposed to classify the impurities as "shallow" and "deep" based on the degree of impurity potential localization [6]. Of course, this classification scheme was introduced with respect to the well-established theory for "shallow" impurities, primarily, those typically referred to as donors and acceptors that were generally believed to have a screened Coulombic impurity potential that was much more extended than the impurity potential of the isoelectronic impurity. In the new framework of the impurity model to be described in this paper, the distinction of the "shallow" and "deep" impurity will essentially disappear, at least on the qualitative level.

## 2. A brief history of the impurity theories

The best known and most widely used theory for a donor or an acceptor like impurity is the so-called hydrogen model with a screened Coulomb potential and an effective mass respectively for either electron or hole. In this model, for a non-degenerate conduction band with parabolic dispersion near a special k point, the donor binding energy $E_D > 0$ is the solution of the equation below [7]:

$$\left(-\frac{\hbar^2}{2m_e}\frac{\partial^2}{\partial r^2} - \frac{e^2}{\epsilon r}\right) F(r) = -E_D F(r) , \qquad (1)$$

where $m_e$ is the electron effective mass, ε is the static dielectric function, and F(r) is the envelope function. The acceptor binding energy $E_A$ can be obtained by replacing $m_e$ with the hole effective mass $m_h$. It is generally believed that except for the region very close to the impurity site where some correction may be needed [2], the screened Coulomb potential should be quite accurate for describing the electron motion away from the impurity site. This simple impurity model actually originated from the early study of anion vacancies in ionic crystals, such as a Cl vacancy in NaCl. The ideal perhaps first appeared in a 1938 paper by Gurney and Mott, where it was suggested that at large distance r from the vacancy the trapped electron would experience an electrostatic field $e/(Kr)$ where K was the dielectric constant, and therefore there would be a series of bound states leading up to a series limit [1]. This idea was implemented in 1939 by Tibbs [2] as a hydrogen mode with the free-electron mass replaced by an electron effective mass of the conduction band, and the vacuum level by the CBM. The model was further discussed by Mott and Gurney in their 1940 classic book entitled "Electronic Processes in Ionic Crystals" [8]. Fig. 1 is an illustration of this model, showing both the potential energy near the defect and the electron wave functions of the defect states [8]. The first application of the hydrogen model to a covalent semiconductor, namely Si, was done by Bethe in 1942 to treat donors in Si [9]. Thereafter, Kittel and Mitchell (1954) [10] and Luttinger and Kohn (1955) [11] extended the single component hydrogen equation to semiconductors like Si and Ge with multiple equivalent extrema in the conduction band (for donors) or degenerated valance bands (for acceptors). One apparent deficiency of the hydrogen model is that the binding energy is independent of atom type, but experimentally the impurity binding energy was found to vary greatly from one impurity to another. For instance, the acceptor binding energy of an acceptor in Si from the hydrogen model is 24.8 meV [12], but experimentally from B to Tl, the binding energy changes from 45.8 to 247.7 meV [13]. The discrepancy between the hydrogen model and the experiment has generally been referred to as a "chemical shift", which is thought to reflect the chemical nature of a specific impurity. Various schemes have been introduced to correct the discrepancy, known as a "central-cell correction", but the concept of the chemical shift or central-cell correction is ill defined and ambiguous. Pantalides and Sah indicated that the inaccuracy of the traditional hydrogen model was caused by the inaccuracy of the impurity potential, i.e., the deviation from the screened Coulomb potential. They developed an "extended effective theory" that was able to improve significantly the accuracy, in particular for isocoric impurities (i.e., impurities in the same row as the host atom) [14]. We will see later that the effective mass theory can be used only for treating one part of the overall impurity problem [4]. However, the effective mass theory is very useful for a wide range of other applications involving a slow varying potential.

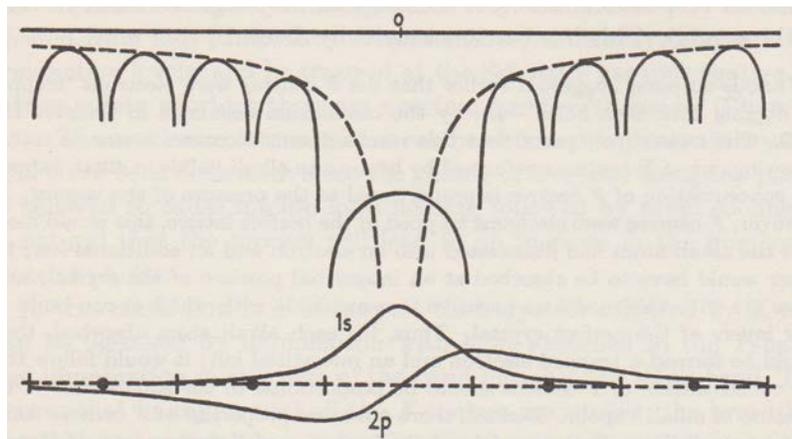

**Fig. 1** Hydrogen model for a Cl vacancy in NaCl given by Tibbs (1939) [2]. Above: potential energy of an electron in the field of a vacant Cl lattice point (full line). The broken line represents $-e^2/(\varepsilon_0 r)$. Below: the envelope wave functions of an electron in a Cl vacancy. + for $Na^+$ ions, ● for $Cl^-$ ions. Source: Mott and Gurney [8].

Today, with the vast improvement in both computation power and theoretical methodology, in principle, we should be able to examine more closely various concepts and models proposed intuitively in the early stage of the semiconductor research, using first-principles based techniques. Indeed, recent advances in first-principles density functional theory (DFT) have made it the tool of choice for studying the properties of defects in semiconductors [15, 16]. However, when coming to compare the computational results with experimental data, one will find it not at all a straightforward task. At the end, a correct conceptual understanding is important in interpreting the first-principles results.

## 3. Electronic structures of donors

At first glance, the model proposed by Gurney and Mott, and Tibbs is quite reasonable. It was stated by Tibbs in his paper [2]: "*Suppose that a negative ion is removed from the interior of such a crystal, leaving a vacant, negative-ion lattice point. This is equivalent to putting a positive charge at the point in the crystal from which the negative ion is removed. … the potential field in the crystal due to this positive charge is e/K.r, where K is the dielectric constant for static fields …*". Tibbs then treated the electronic states in the field of the positive charge using an effective-mass hydrogen model. Note that the crystal with a missing ion, a charged system, usually is unstable, thus will be neutralized by an electron in the environment. As stated by Tibbs, "if we introduce an electron into the lowest state of this potential hole the crystal is again electrically neutral". Actually, for the crystal to remain charge neutral, the "added-back" electron does not have to go into the lowest state but can be anywhere in the crystal. Therefore, it makes more sense to consider an alternative scheme, i.e., removing one (neutral) Cl atom as a whole. In this way, the crystal always remains charge neutral. One could view the system with such a vacancy to be equivalent to a system with a substitutional impurity of a virtual "empty" atom, i.e., a small vacuum space, replacing the Cl atom. Either way, in real space, one will need to decide where to put the "added-back" electron or the extra valence electron of the nearby Na atoms that would have been transferred to the removed Cl atom. In energy space, the question will be: which energy level should this electron go into? There are two distinct options of allocating the extra electron: one is to place it back to the vacancy site; another is to let it be away from the defect site, although it might still sense the attractive force of the potential hole. For the latter option, if the electron is able to escape the attraction all together, it will become a free conducting electron in the conduction band. The situation is more complicated for the former option. One would need to specify an energy state with its wave function being largely localized at the vacancy site. What kind of state is it expected to be and where should its energy level be? Let us do a simple electron counting. Assuming that the defect-free crystal has $N_0$ occupied valence states for hosting $2N_0$ valence electrons, the defected crystal has $N_0 - 4$ valence states due to missing one Cl atom (one 3s orbital, and three 3p orbitals), and can accommodate $2N_0 - 8$ valence electrons. In the meantime, the system has $2N_0 - 7$ valence electrons, which implies that one valence electron (the one from the Na atoms next to the vacancy) will need to occupy an excited state above the fully occupied valence band. One of the occupied valence states is in fact a singlet bonding-state of the vacancy, which could be understood as a bonding state of the Na "dangling bonds". This bonding state is expected to lie deeply in the valence band, known as a "hyper-deep" defect state, which is localized in the vicinity of but not exactly at the vacancy site, because the vacancy site (vacuum) has the highest electron potential energy. There will be correspondingly a singlet anti-bonding state somewhere close to the conduction band [3]. The defect state that can host the orphan electron is exactly this anti-bonding state that is expected to be more localized at the defect site than the "hyper-deep" bonding state. With this physical picture in mind, one may now realize that the exact position of this energy level should depend on the difference between the vacuum level and the atomic orbitals of the atoms involved. The question to ask would be: is it at all reasonable to expect this defect state to be like a hydrogen ground state? One could already see that the formation of this defect state is quite complicated such that the possibility for it to be a hydrogen like state is rather small.

Is there any problem with the idea that the Cl vacancy in NaCl or more general a donor atom in a semiconductor (e.g., P in Si) would generate a hydrogen like long-range potential away from the defect site? For the case of Cl vacancy, if the extra electron is taken away from the defect site, there will be indeed a positive charge at the vacancy site with respect to the defect free ionic crystal. For a hydrogen atom, whether or not the electron is present, the Coulomb potential generated by the core remains the same, i.e., $-e/r$, and the electron energy levels are determined by this potential. However, for a more complex atom, even He, the single particle potential for a He ion ($He^+$) is significantly different from that of $H^+$. Therefore, even though both $H^+$ and $He^+$ have the same positive charge, the ionization energies of H and He are very different (13.6 eV vs. 24.6 eV). Based on this consideration alone, one cannot really expect that the electronic structure of a donor would be something close to what predicted by the hydrogen model, because of the many-particle effect and/or the variation in the detailed bonding situation with the host. With this understanding, we can say that the model potential depicted in Fig. 1 is most likely overly simplified for calculating the ionization energy of the neutral vacancy state. As a matter of fact, in the typical DFT description of a semiconductor, the atomic potentials are usually highly localized, and therefore in the ground state the effective single particle potential does not have any long-range component that would extend much beyond the defect site. However, it is also well-known that in the DFT level defect calculation, the effective single particle potential for the excited state (e.g., $V^+$) tends to have a long-range Coulombic component. In fact, it is the existence of such long range interaction that requires corrections to remove the spurious effects in the supercell based defect calculation due to the use of a finite supercell size. These qualitative understandings have been confirmed by a DFT modeling of the Cl defect in NaCl, where a single defect state (corresponding to the anti-bonding defect state in the above discussion) is found to be highly localized at the vacancy site in the neutral state $V^0$ but less so in the ionized state $V^+$ [17].

To help readers visualize the ground state electron distribution in NaCl with a vacancy, a one-dimensional (1-D) model is provided in Fig. 2 to illustrate the electron occupation in the spirit of a tight-binding model. Although the above discussions were for a specific donor-like defect, the general consideration and argument may apply to other donor-like point defects and impurities. One noticeable qualitative difference between the donor-like vacancy and a donor atom (e.g., P in Si) lies in that for the latter the occupied impurity level of the neutral donor state (equivalent to the one in Fig. 2) would be a bonding state in nature.

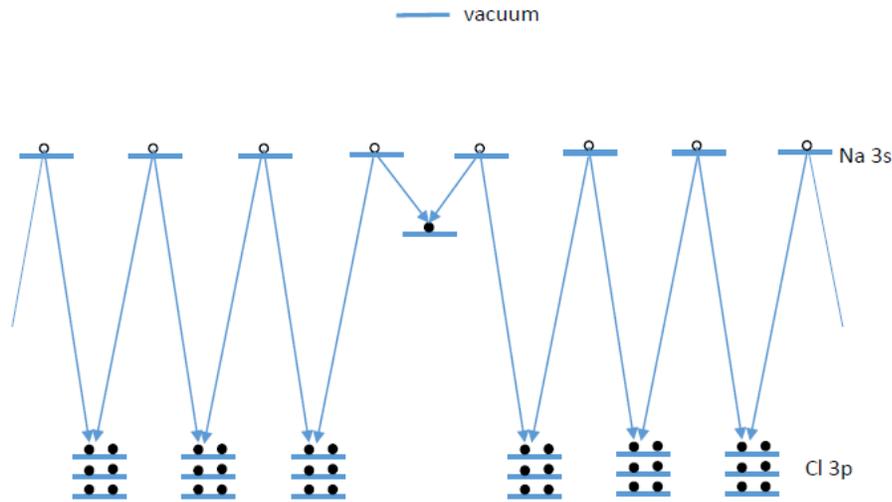

**Fig. 2** 1-D electron occupation model for NaCl crystal with one Cl vacancy for the ground state of the system or the neutral vacancy state $V^0$.

Now let us accept that a Cl vacancy does introduce a bound state below the CBM with a binding energy $E_I$, which will be referred to as impurity binding energy. In the single electron picture, the conduction band states represent the electronic states free from the interaction with the defect center. However, when the electron tries to escape from the defect site or is excited from the lowest defect level, a Coulombic attractive potential arises, which qualitatively corresponds to the situation described in Fig. 1. The strength of this attractive potential only provides some extra holding-back force preventing the electron from escaping from the defect, i.e., going into the conduction band. This problem is much like the exciton effect in a normal semiconductor, if we view the defect state as part of the ground state of the system. To make this even more like a standard exciton problem, we may envision that there is an array of, but nevertheless well separated, defects in the crystal such that the interaction among the defects is very weak but we could still think of having a new valence band that is separated from the conduction band with a bandgap approximately given by $E_I$. In this picture, a relatively small exciton binding energy, compared to the electron binding energy, will emerge naturally from the attractive Coulomb potential in the same manner as in the standard exciton problem. For an isolated defect center, we will have a bound exciton with a hole at the defect site and an electron orbiting around the defect, and for this bound exciton problem the dielectric screening and electron effective mass can be more justifiably used. This is basically the bound exciton model that unifies the "shallow" and "deep" impurities and defects [4]. The exciton binding energy (referred to as $E_D$ for a donor like defect) in fact resembles what is normally considered as the donor binding energy $E_D$ given by the solution of Eq. (1), which is obviously very different from the defect binding energy $E_I$ introduced above. If an electron is bound to the defect, the energy needed to set it free in the conduction band or to induce electron conductivity will be the defect binding energy $E_I$ instead of the exciton binding energy $E_D$ that is given by the popular hydrogen model. Fig. 3 compares the energy band diagrams of a Cl vacancy like donor defect in the conventional and revised picture. In a nutshell, on the conceptual level, the defect problem is very much similar to the exciton problem in a semiconductor, although the "valence band" (the neural defect state) in this particular example is only half occupied. What is the deficiency of the conventional model? In the conventional picture, as shown in Fig. 3(a), the 1s level of the neutral donor state is described by a hydrogen model that also gives rises to many excited states 2s, 3s, … If only for the concern of the magnitude of the binding energy, this picture could be viewed as an oversimplification of treating a many-electron atom with an one-electron atom. Then, it might be reasonable to correct this shortcoming with some sort of "core correction" to get the correct impurity binding energy $E_I$, as depicted in Fig. 3(b). However, in a real self-consistent many-electron calculation, after carrying out this correction, one might or might not get those excited states shown in Fig. 3(a). For instance, in the ground state calculation, i.e., the electron stays with the donor, the impurity potential tends to be highly localized, thus, might not generate an excited state. This issue will be discussed later in the first-principles calculation section. On the conceptual level, the process described in Fig. 3(c), the bound exciton aspect, does not exist in the conventional model. Clearly, the electronic transitions in Fig. 3(a) and Fig. 3(c) are different: for instance, in the former case, the first electronic transition would be the 1s to 2p transition, whereas in the latter case it is the neutral donor to 1s bound exciton transition. The bound exciton concept addresses the correlation effect of the excited electron with the remaining valence electrons, including all of those either from the donor atom or host atoms. The hydrogen model turns out to be relevant only to part of the bound exciton problem.

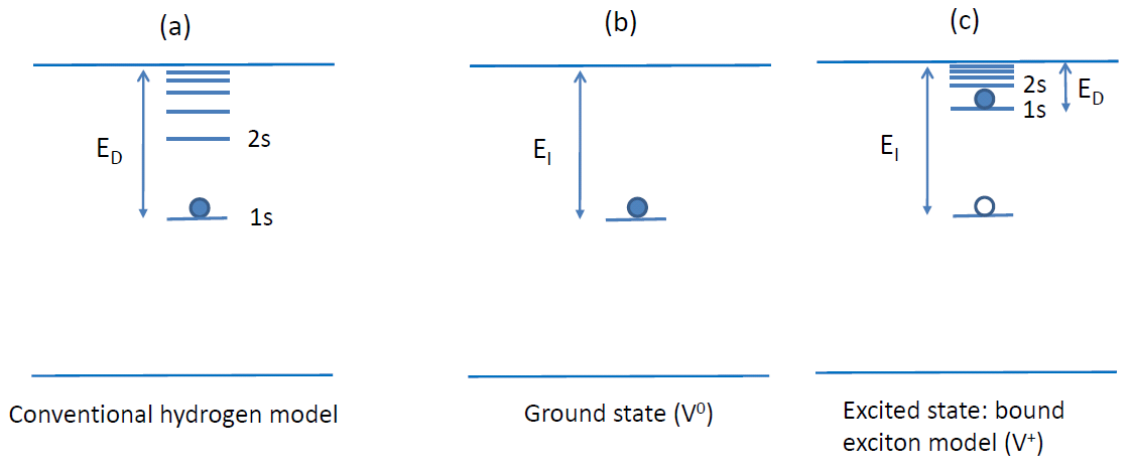

**Fig. 3** Energy band diagrams for a Cl vacancy like donor. (a) Conventional hydrogen model for a donor in its ground state with a binding energy $E_D$. (b) Neutral vacancy state ($V^0$) with an electron binding energy $E_I$, and (c) the lowest excited state of the vacancy (one of the possible $V^-$ states) – a bound exciton with a binding energy $E_D$.

We will next discuss acceptor-like impurities to further help the conceptual understanding offered above. The case of the acceptor seems to be somewhat more transparent in physics than that of the donor.

**4. Electronic structures of acceptors**

When an impurity with one or more valence electron(s) less than that of the replaced host atom is introduced into an otherwise perfect semiconductor, it typically introduces a partially occupied state near the top of the valence band. Such an impurity is often referred to as an acceptor, because it can accept one or more electron(s) from the valence band by thermal excitation, assuming these states are relatively close to the VBM. Here we do not consider the trivial case where the impurity level turns out to be below the VBM, and thus the acceptor will be self-ionized (i.e., generating a hole in the valence band), effectively resulting in a metallic material. In a typical textbook description, an acceptor is negative charged center that has an attractive Coulomb potential for the hole, which introduces a hole bound state or an empty electron level at energy $E_A$ above the VBM. $E_A$ is known as acceptor binding energy, and understood as the energy needed to promote an electron from the VBM to the impurity level and thus generate a free hole in the valence band. Consequently, the transition energy for an electron in the conduction band to the acceptor level, known as a free-to-bound transition, would be $E_{F-B} = E_g - E_A$, where $E_g$ is the bandgap. With this understanding, the energy diagram of an acceptor center and the related transition energies are illustrated in Fig. 4(a), as appeared virtually in all textbooks. The corresponding solutions of Eq. (1), i.e., those energy levels above the VBM as shown in Fig. 4(a), are meant to be the hole bound states. However, the physical meaning of those states are not straightforward. One may interpret exciting an electron from the valence band to the 1s hole state as generating a free hole in the valence band, but the meaning of exciting one electron from the valence band to the 2s, 2p, … hole states becomes ambiguous. More explicitly, for instance, if an electron were excited to the 2s hole state, the transition would not generate a free hole in the valence band, because the excitation energy is short by an amount of $E_{1s} - E_{2s}$ for generating a free hole. One would have to think of having a hole at an energy $E_{1s} - E_{2s}$ above the VBM. Typically, in the conventional acceptor model, the optical transition is understood solely as that of the hole making transition from its 1s state to different p-like excited states.[18] However, as we know, the transition of a hole is just a convenient way of understanding an electronic transition. Although one could understand the transition between the VBM and 1s state in Fig. 4(a) as either an electron transition from the VBM to the 1s acceptor state or a hole transition from the 1s acceptor state to the VBM, the hole transition from 1s to 2s, 2p, … cannot be easily associated with equivalent electron transitions. Basically, these 2s, 2p, … hole excited states cannot be simply understood as unoccupied electronic states to which an electron can make a transition. This awkward situation will disappear in the new model, as will be explained below.

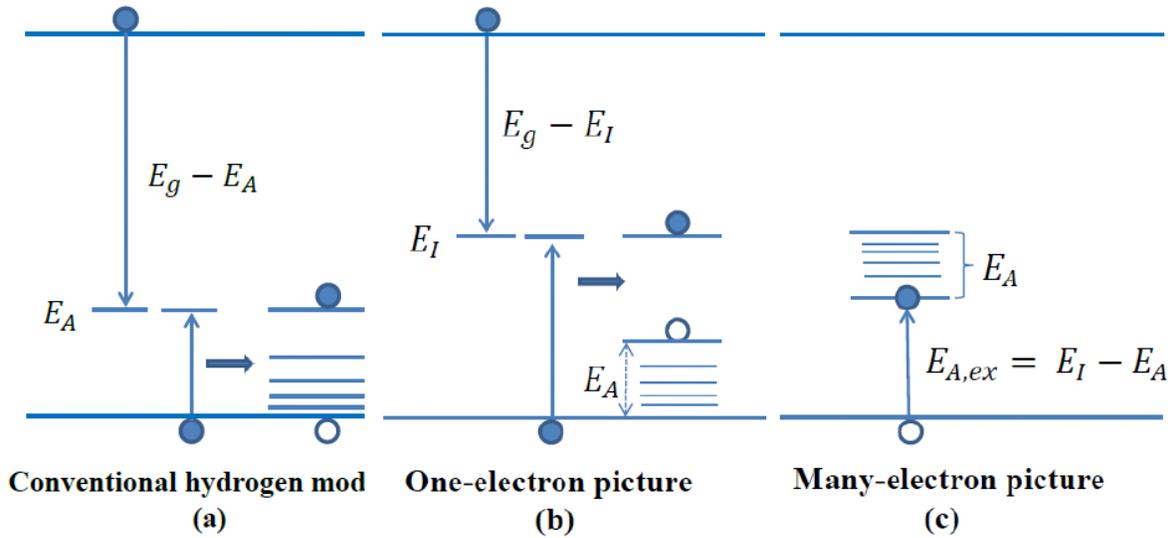

**Fig. 4** Energy band diagrams for an acceptor. (a) Conventional hydrogen model for an acceptor. When an electron is excited into the hole bound state at $E_A$, a free hole is generated in the valence band. (b) Bound exciton model for an acceptor. When an electron is excited into the impurity state at $E_I$, a bound exciton is formed with a hole binding energy $E_A$. (c) The bound exciton model in a many-electron picture.

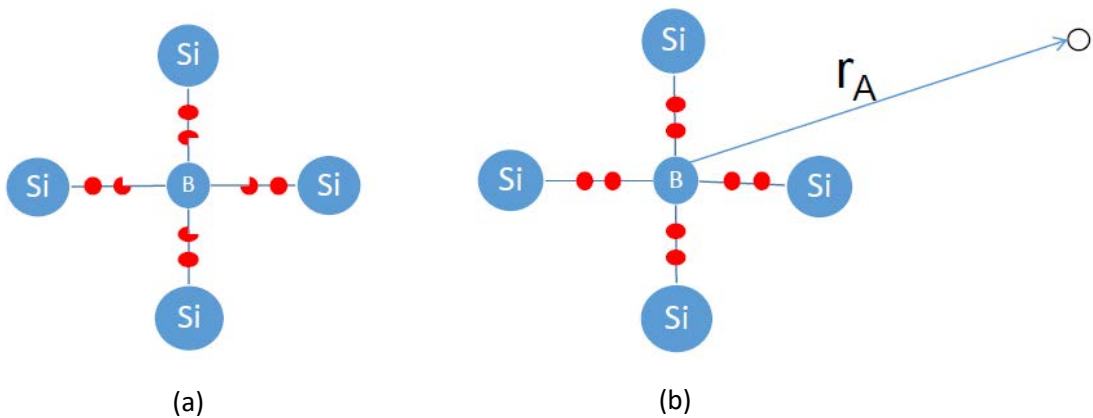

**Fig. 5** Boron acceptor in Si crystal. (a) Neutral or ground state of the acceptor with approximately 1.75 electrons on each bond. (b) Charged or ionized state of the acceptor with approximately 2 electrons on each bond, and a hole bound to the negatively charged acceptor core.

Again the above mentioned standard textbook description about acceptor is conceptually problematic. Despite having less valence electrons, the acceptor impurity is in fact charge neutral if the sample temperature is sufficiently low. It is important to understand that a neutral acceptor impurity does not have a long range Coulomb potential centered at the impurity site. The attractive Coulomb potential in Eq.(1) only arises after one electron has "jumped" into the acceptor site. This process involves an electronic transition of an electron being be excited from the valence band into a higher energy level that is provided by the acceptor impurity. The electronic transition is schematically shown in Fig. 5, where Fig. 5(a) represents the neutral or ground state of the acceptor with one less valence electron on the four bonds between the nearest neighbor Si and the impurity, and Fig. 5(b) illustrates that one electron has been excited into the empty impurity state or brought to the impurity site from the adjacent area, resulting in one hole bound to the negatively charged impurity center through Coulomb interaction. We assume that this impurity state is at $E_I$ above the VBM, and $E_I$ is referred to as impurity binding energy, similar to the case of the donor, as illustrated in Fig. 4(b). Then, what determines $E_I$? Let us first consider qualitatively why group III elements B, Al, Ga, In, Tl will create acceptor states in Si, and why their binding energies increases in the order of going down the column. The hints are in the energy diagram of the valence states for those most interested elements for semiconductors, as shown in Fig. 6. Fig. 6(a) depicts the energy levels of the valence electrons for most elements found in the commonly encountered semiconductors, whereas Fig. 6(b) highlights the p orbital energies of the group III elements with respect to that of the Si p orbital together with the experimentally determined impurity binding energies for these acceptor impurities. Firstly, we recognize that the Si valence band is primarily derived from the 3p orbitals of Si atoms. Secondly, Fig. 6(a) shows that the 2p states of B, 3p states of Al, …, and 6p states of Tl are all higher than the Si 3p states. Therefore, they all have the tendency to form p-like impurity states above the Si VBM, and they all actual do, with the binding energy increasing in the same order, as also shown in Fig. 6(b). This observation suggests that the value of $E_I$ is closely related to the difference in the atomic orbitals between the host and impurity atom, thus obviously should depend sensitively on the impurity species. Because the variation in the np atomic states are the consequence of the many-electron effect in the atomic structure, which results in the major deviation from that of the hydrogen atom, there is no obvious reason to think that the impurity state can be predicted by a hydrogen like model. Quite naturally, we expect that the acceptor level position should vary from one atom to another.

Till now we have implicitly assumed that the acceptor level to be occupied by the electron taken from the valence band is related to the p valence state or the first ionization energy of the acceptor atom. Conceptually, because we are adding an additional electron to the impurity atom, this level should resemble more of the second ionization level or correlate with the electron affinity of the impurity atom. In a free standing atom, this level tends to be much higher than the valence state due to the screening effect of the core electron(s). In a crystal, because of the delocalization of the valence electrons, the screening effect is expected to be much weaker, so we may not need to emphasize this subtle issue at least in the qualitative level. However, for B its 2p level is very close to the Si 3p. Although shown to be somewhat higher in Fig. 6(b) according to one calculation [19], it is actually slightly lower based on other literature values of their ionization energies (8.30 eV for B and 8.15 eV for Si), which means that simply based on the ionization energy consideration, B might not form a bound state in Si. On the other hand, B has a small electron affinity because of the strong screening effect, in fact smaller than Al (0.28 eV for B vs. 0.43 eV for Al), i.e., it is more likely to form a bound state than Al in Si if the electron affinity is concern, which could explain qualitatively why B is able to form a bound state in Si.

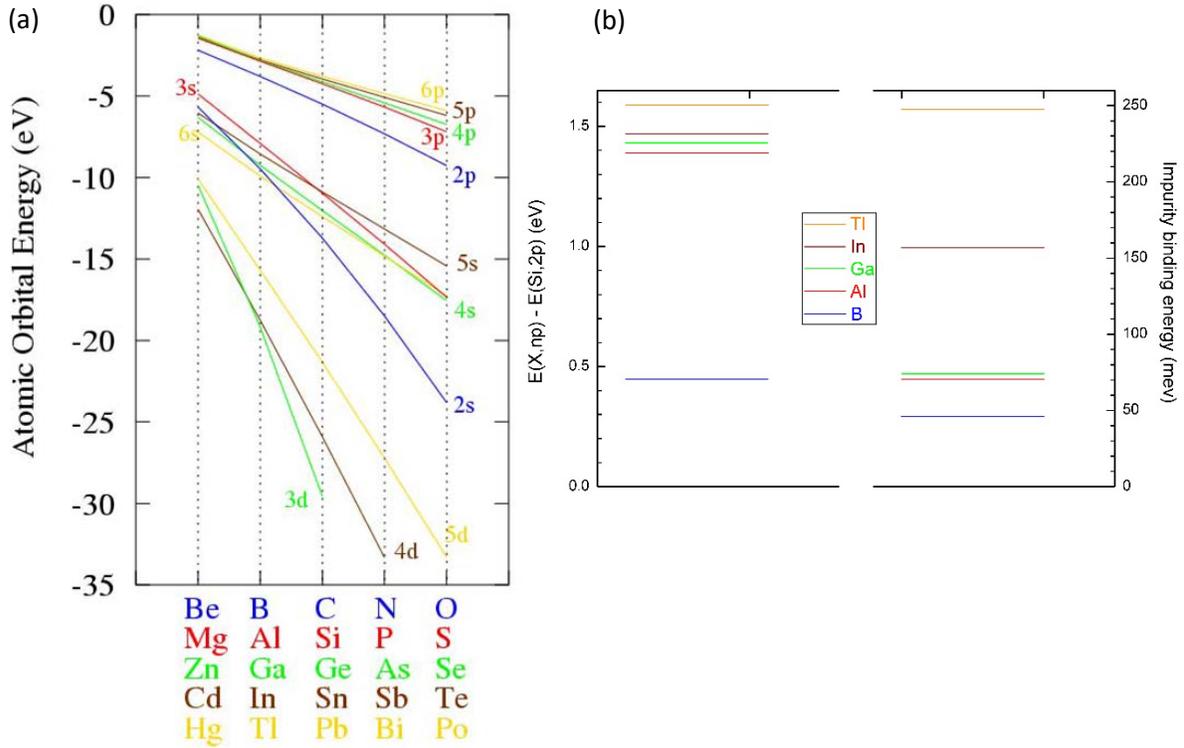

**Fig. 6** Comparison of valence atomic energy levels with the acceptor impurity binding energies in Si. (a) Valence atomic energy levels calculated using a density functional theory within a local density approximation for most group II – VI elements (provided by Suhuai Wei [19]). (b) Left: p valence electron energy levels of Group III elements with respect to that of Si 3p energy level (from the graph in (a)); Right: experimental values of the acceptor impurity binding energies of Group III elements in Si.

In the next step, after the acceptor level is occupied by an electron from the valence band, the hole left behind is not free but still attracted to the ionized center $A^-$. It is this potential that give rises to those hole bound states shown in Fig. 4(b) or binds a hole as illustrated by Fig. 5(b).

If the impurity state is very tightly localized at the acceptor site, the A⁻ center can be viewed as having a point charge –e, the bound states of the hole will be the solutions of Eq. (1) or a set of generalized envelope function equations taking into account the complexity of the real valence band [12, 20]. This is again an exciton bound to the impurity center with a hole binding energy $E_A$, with a series of excited states for the exciton, as illustrated in Fig. 4(c). Ideally, if the point charge assumption is valid, $E_A$ will be independent of the impurity species. In this bound exciton state, the hole is not free, thus, an extra energy $E_A$ is required to set the hole free, which means that the energy required to create a free hole in the valence band is $E_I$, just as in the case of the donor. A tight-binding model can make this picture easier to understand. Let us consider an acceptor impurity, with one less valence electron, having a higher p-orbital than the host. If one electron is moved from the host to the impurity site by applying an excitation energy of the difference between the p-orbitals, a hole is then generated in the valence band, but remains attracted to the ionized impurity through the Coulomb interaction. Additional energy is needed to allow this bound hole to break away from the acceptor becoming a free hole. Compared to the discussions given earlier for the case of a donor, the physics picture for an acceptor seems to be somewhat straightforward, a clearer resemblance with the free exciton problem in a semiconductor.

One complication should be noted, that is, an acceptor typically introduces multiple impurity states of which are either fully or partially occupied, instead of merely one empty state. For instance, an element with three p electrons replaces one host atom in Si, the impurity states will be p-like occupied by five p-like valence electrons (of which 4 from the nearby Si atoms and one from the impurity) with one empty state. If a spin-orbit interaction is taken into account, these p-like impurity states will split into two states with the lower one fully occupied. When drawing a band diagram, one typically ignores the existence of occupied impurity states.

What exactly is the difference between the above described acceptor model and the conventional one? It is interesting to read the description about the acceptor in a classic book (published in 1950) entitled "Electrons and Holes in Semiconductors" by Shockley,[21] which is quoted here: "*The hole in one of the bonds to the boron atom can be filled by an electron from an adjacent bond, and the hole can thus migrate away, as described in Figure 1.5 (d). The boron thus becomes an immobile, localized negative charge. Because of the symmetry between the behavior of holes and electrons, we can describe the situation shown in Figure 1.8 by saying that the negative boron atom attracts the positively charged hole but that thermal agitation shakes the latter off at room temperature so that it is free to wander about and contribute to the conductivity.*" Although the conventional model does recognize the transfer of the additional electron from the adjacent Si, it is implicitly assumed the transfer occurs spontaneously without costing any energy. Therefore, the conventional acceptor model somehow skips the first step or fails to recognize the independent identity of the neutral impurity state. There, this state is either non-existed or implied to be the same as the hole bound state with a binding energy $E_A$. After understanding this subtle point, we may conclude that the acceptor binding energy described by the hydrogen model in the conventional acceptor theory is in fact equivalent to the hole binding energy in the bound exciton model for the acceptor. The underlying physics of the bound exciton model can be further clarified after the acceptor bound exciton is compared with an isoelectronic impurity bound exciton in the next section.

## 5. Unification of the "shallow" and "deep" impurities

We first describe the electronic structure of an exciton bound to an isoelectronic impurity of an electron trap, such as GaP:N, where the bound exciton is known as an "acceptor-like bound exciton" based on the model proposed by Hopfield, Thomas, and Lynch (HTL model) [22, 23].

This bound exciton problem is viewed as a classic example of the "deep" impurity that is thought to be profoundly different from the "shallow" impurity, either an acceptor or a donor, in terms of the extension of the impurity potential [6]. As illustrated in Fig. 7(a), for an isoelectronic impurity N in GaP, the nitrogen atom generates an electron bound state within the bandgap with a binding energy $E_e$ with respect to the CBM or an impurity state $E_N$ measured from the VBM. Being an empty state far away from the VBM, this impurity state can be considered as a deep acceptor level. In the one-electron picture, the formation of a bound exciton on the N impurity could be viewed as a two-step process (HTL model): (1) one electron is excited into the electron bound state from the valence band (or captured from the conduction band if the electron was already in the conduction band), forming a so-called bare electron bound state or a negatively charged center $N^-$; (2) through the Coulomb interaction, a hole is attracted to the $N^-$ center, forming a bound exciton with a hole binding energy $E_h$ (with respect to a hole at the VBM), as shown in Fig. 7(a), where the transition energy $E_{N,ex} = E_g - (E_e + E_h) = E_N - E_h$ corresponds to the zero-phonon absorption or emission energy of the bound exciton. Note that the term "acceptor-like bound exciton" is not referring to the empty impurity bound state that is acceptor like, but the similarity between the process of the hole bound to the $N^-$ center and the hole bound to an acceptor that was viewed to be a negative charged center according to its conventional model [22, 23]. With the new understanding, the N isoelectronic center in GaP and B impurity in Si are qualitatively the same: both are acceptors and can form a bound exciton state. However, there are some subtle differences: for the former the bare electron bound state is s-like anti-bonding state, and for the latter the bare electron bound states are p-like bonding states; the former is an empty state, and the latter partially occupied. It was largely because of the misunderstanding about the conventional donor and acceptor impurities, the "deep" impurity concept [6] was proposed to emphasize the highly localized nature of some electronic bound states like N in GaP and GaAs to contrast with the "shallow" impurities that were perceived to have a long-range Coulomb potential. Now the distinction has practically vanished with the unified understanding given above. We can also see the similarity between the "donor-like bound exciton" for GaP:Bi [22] and a simple donor center. In the case of a Bi impurity in GaP, Bi introduces p-like bound states above the GaP VBM, because Bi 6p states are sufficiently higher than P 3p states, though not high enough than As 4p to also form the bound states in GaAs:Bi [24]. In GaP:Bi, when a bound electron is excited by a photon with an appropriate energy to reach the conduction band, an bound exciton can be formed at the Bi site [25], because of the creation of a $Bi^+$ center just like the case of the $V^+$ state for the Cl vacancy in NaCl.

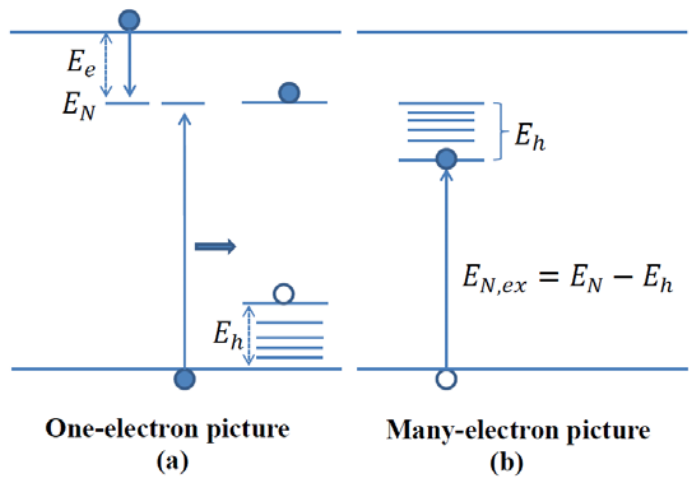

**Fig. 7** Energy band diagrams for an exciton bound to an isoelectronic impurity N in GaP, where $E_N$ is the energy level of the bare electron bound state measured from the VBM, $E_h$ the hole binding energy to the $N^-$ center, and $E_{N,ex} = E_N - E_h$ the lowest bound exciton transition energy. (a) In one electron picture, and (b) in many-electron picture.

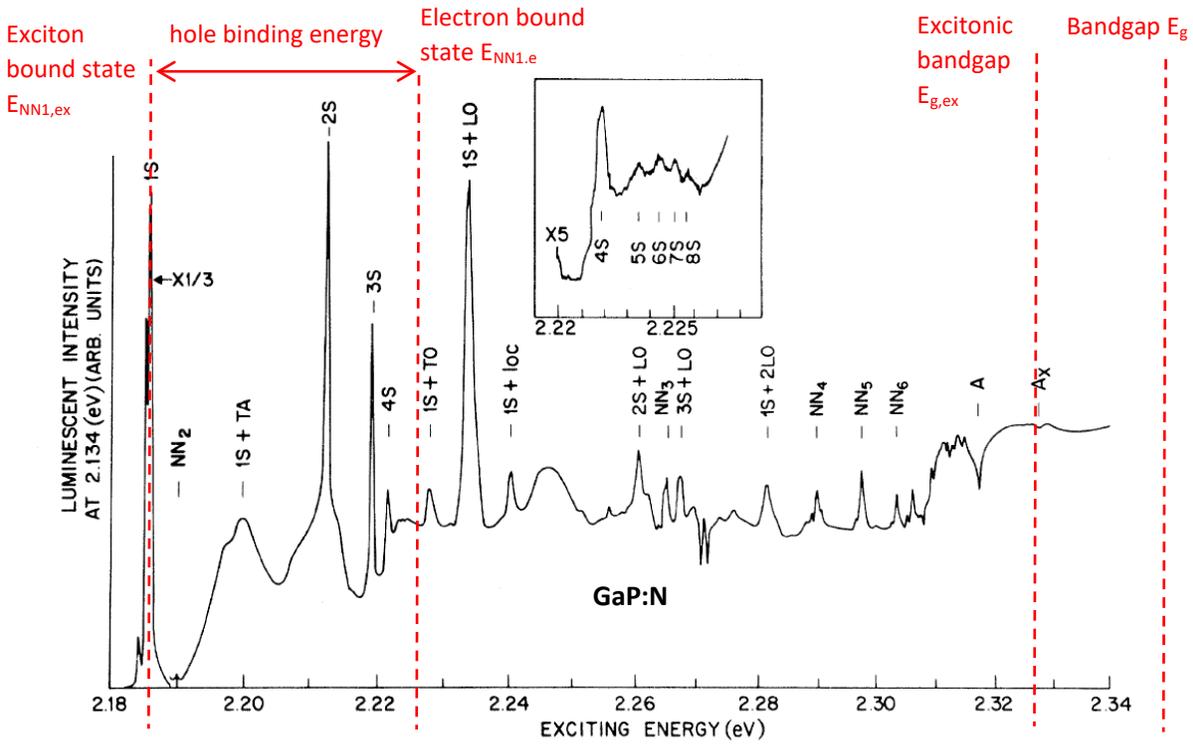

**Fig. 8** Photoluminescence excitation spectrum of $NN_1$ center in GaP:N (from Cohen and Sturge [26]). Vertical dashed lines and labels in red color are added to indicate the primary transition energies.

We now explicitly compare the absorption spectra of an isoelectronic bound exciton and an acceptor bound exciton to illustrate the similarity in their electronic structures. Fig. 7(a) shows a series of hydrogen-like excited states for the hole bound to the N⁻ center in GaP:N. Such excited states have not been unambiguously observed experimentally for the isolated N center, but have been observed for many N-N pair centers $NN_i$ by Cohen and Sturge [26]. Shown as an example in Fig. 8 is a photoluminescence excitation spectrum (very similar to an absorption spectrum) for the deepest N-N pair $NN_1$ that has an electron binding energy $E_e = 125$ meV, and hole binding energy $E_h = 40.3$ meV. Furthermore, $E_h$ was found to decrease with reducing $E_e$, for instance, down to ~34 meV when $E_e = 10$ meV for $NN_7$. $E_h$ values were systematically smaller than the "acceptor binding energy" calculated using the effective mass theory $E_A$ = 47.1 meV [12, 26]. Initially the deviation was explained as due to a repulsive core correction that was negative in this case (in contrast to the normal case being positive) [27], but later was pointed out as due to the finite extension of the electron bound state [28]. If the electron bound state were perfectly localized at the impurity center like a point charge, $E_h$ would approach $E_A$, which explains the observed dependence of $E_h$ on $E_e$ among different $NN_i$ centers. The observation of the nS series of bound exciton excited states was considered to be the most solid evidence for the validity of the HTL model [22]. However, the community of the isoelectronic impurity study did not recognize the potential impact of this understanding to the well accepted model for the so called "shallow impurities", because the validity of the general understanding for the "shallow impurities" was taken for granted. Therefore, it was understandable to refer these bound excitons as "acceptor-like", because based on the understanding of the time there was indeed a significant difference: no impurity bound state corresponding to the $E_N$ level in the standard model for the acceptor.

The bound exciton formation is ultimately a many-electron problem that should be treated as the transition between two states of the whole system. Fig. 7(b) shows the energy diagram in the many-electron picture, where $E_N = E_g - E_e$ can be viewed as the upper limit of the bound exciton states corresponding to the hole in different hydrogen-like bound states 1s, 2s, etc.. This many-electron picture makes it easier to understand why the formation of a bare electron state is not a necessary pre-cursor to the formation of a bound exciton. Rather a bound exciton can be formed with a resonant excitation of energy $E_{N,ex}$. The relationship between $E_N$ and $E_{N,ex} = E_N - E_h$ is analogous to that between $E_g$ and the free exciton bandgap $E_{g,ex} = E_g - E_{ex,b}$ in a semiconductor, and $E_h$ corresponds to the free exciton binding energy $E_{ex,b}$ (e.g., $E_{ex,b}$ = 22 meV for GaP, and 20.6 meV for Si). Different excitonic states represent different degrees of electron-hole correlation.

When the correlation vanishes, the hole is free. Similarly, in the many-electron picture, the bound exciton model for the acceptor and donor are given in Fig. 4(c) and Fig. 3(c), respectively.

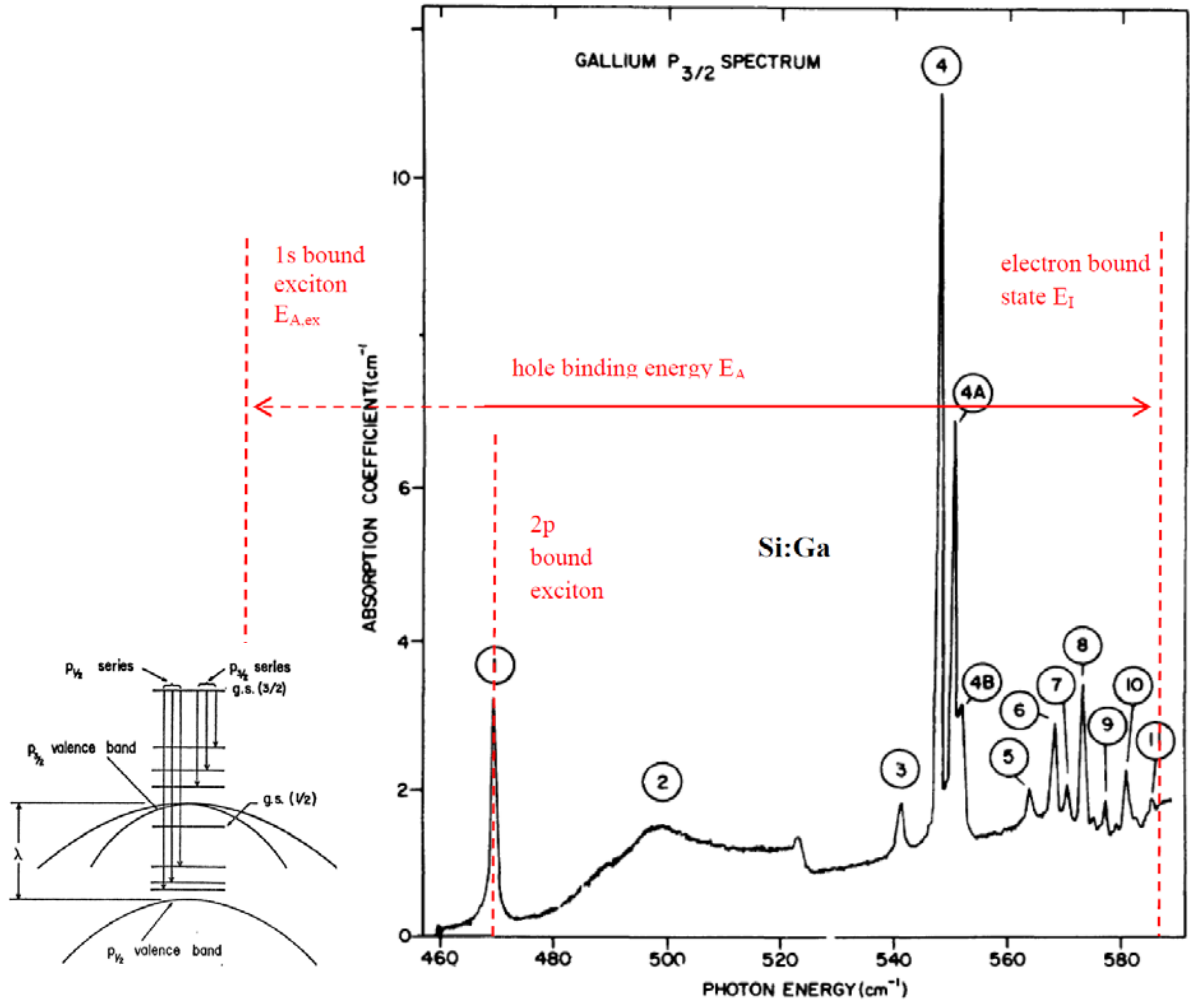

**Fig. 9** Absorption spectrum of Ga in Si (from Fischer and Rome [29]). The inset shows the transitions between hole bound states (from Onton, Fisher, and Ramdas [18]). Vertical dashed lines and labels in red color are added to indicate the primary transition energies.

We now compare the absorption spectrum of an acceptor with the example of the isoelectronic bound exciton. Fig. 9 shows an IR absorption spectrum for Si:Ga [29], where the lowest energy transition (labeled as "1") is assigned as 1s-to-2p like transition between the hole bound states in the conventional hydrogen model, as shown in the inset [18]. In the new understanding, the lowest transition will correspond to the transition from the ground state of the system to the 2p like bound exciton state, as illustrated in Fig. 4(c), and those discrete transition lines will approach the limit of impurity bound state $E_I$. Note that for the acceptor, the VBM to 1s bound exciton transition is dipole forbidden, and first allowed optical transition is the ground state to 2p bound exciton state, because the electron bound state is mostly p-like, whereas for the nitrogen bound exciton, the ground state to the 1s bound exciton state is dipole allowed, because the electron bound state is mostly s-like. The similarity (as well as the subtle difference) between the acceptor (more accurately acceptor bound exciton), isoelectronic bound exciton or free exciton has not been recognized in the past. Obviously, with this revised view of the "shallow" impurities, many experimental data in the literature need to be re-interpreted.

In many cases, different interpretations might not have any practical consequence. For instance, to generate free holes in a p-type semiconductor, in the conventional model, electrons are thermally excited to the acceptor level with an activation energy of $E_A$, as shown in Fig. 4(a); whereas in the bound exciton model, Fig. 4(b), the electrons should be excited into the impurity state $E_I$. In either case, experimentally one would just observe a thermally activated conductivity, regardless what the underlying physical process is. However, in some other cases, the excitonic model allows for easier understanding of the underlying physics. For example, in Zn and N co-doped GaP and at low temperature, an optical excitation with an energy of $E_{NN,ex}$ [the bound exciton energy of a nitrogen pair, equivalent to $E_{N,ex}$ in Fig. 7(b)] was found to yield an inter-impurity transition interpreted as between $E_{NN}$ of the nitrogen pair [the bare electron bond state energy, equivalent to $E_N$ in Fig. 7(a)] and $E_A$ of the Zn acceptor [Fig. 4(a)] [30]. There is one aspect of this experiment that cannot be easily explained within the one electron picture, that is, the excitation photon does not have enough energy to populate the $E_{NN}$ state or an additional energy of $E_h$ is required to release the hole from the NN pair bound exciton to facilitate the $E_{NN}$ to $E_A$ transition. Now in the excitonic picture, the transition can be simply explained as a bound exciton is transferred from a NN center to the Zn center, accompanying by the emission of a photon with an energy equals to $h\nu = E_{NN,ex} - E_{A,ex} = (E_{NN} - E_I) - (E_h - E_A)$. Because $E_h$ and $E_A$ are expected to be close to each other (although they could differ by a few meV), we have $h\nu \approx E_{NN} - E_I$, which justifies the original interpretation. In another example of a recent interest to use a superlattice scheme for improving p-type conductivity in III-nitride devices, it has been pointed out that adopting the conventional or revised acceptor model would mean using different impurity distributions between the barrier and well regions in an optimal design [31].

All examples mentioned above involve relatively simple impurities or defects. For more complex situations, such as transition metal impurities, also known as magnetic impurities, the analyses are less straightforward [32, 33]. Nevertheless, the excitonic effect is expected to also play the similar role in related optical transitions. This effect has so far been neglected in the literature.

## 6. First-principles theories for impurities

Next we discuss how the electronic structure of an impurity can be calculated using first-principles theories. Different schemes have been developed for this purpose, but their results are not necessarily equivalent, not only because they involve different approximations but also because what they calculate can be different things. In virtually all DFT calculations for impurities or point defects, the calculated transition energies were either explicitly or implicitly treated as $E_A$ or were compared to experimental results that have been interpreted as $E_A$. Based on the discussions given in the previous sections, this practice is problematic. We will clarify the differences in terms of the underlying physics between various transition energies associated with an acceptor calculated using different density-functional theory (DFT) based approaches.

We first offer some qualitative discussions based on a Hartree-Fock (H-F) approximation that is also a many-body theory but seems to be conceptually more transparent than a DFT for illustrating the underlying physics. Within the H-F approximation, the total energy difference between the two states of the system, the excited state (one electron has been moved to the impurity state from the VBM) and the ground state (the valence band is fully occupied), is given as:[34]

$$\delta E_{tot} = E_I - E_{VBM} - \left[ \langle \varphi_I \varphi_{VBM} \left| \frac{e^2}{\varepsilon r} \right| \varphi_I \varphi_{VBM} \rangle - \langle \varphi_I \varphi_{VBM} \left| \frac{e^2}{r} \right| \varphi_{VBM} \varphi_I \rangle \right], \qquad (2)$$

where $E_I$ and $E_{VBM}$ are the absolute values of H-F one electron eigen energies for the impurity and VBM state, and $\varphi_I$ and $\varphi_{VBM}$ are the corresponding wave functions, respectively. We should assume $E_{VBM} = 0$ so that the meaning of $E_I$ in Eq.(2) is consistent with the same quantity introduced above. The first term in the square brackets is the Coulomb interaction between the impurity state and VBM, and the second term is the exchange interaction. The dielectric function ε is added empirically to the Coulomb interaction term to include the screening effect, but not to the exchange term because of its short-range nature. From now on, we will refer both Coulomb and exchange interaction together as Coulomb contribution for simplicity. Conceptually, this Coulomb contribution is really what the $E_A$ in Eq.(1) is about, and it occurs only after the transition of one electron from VBM to $E_I$ has occurred. Physically, it represents the net change of the Coulomb interactions among all valence electrons of which one has been promoted to the $E_I$ state. A DFT version of Eq.(2) is given by Eq.(15) of Ref.[35]. If we take the Coulomb term as an approximation for $E_A$ in Eq.(1), the total energy difference will then be $\delta E_{tot} \approx E_{A,ex} = E_I - E_A$. Evidently, the reason for the approximate sign is that $\delta E_{tot}$ given by Eq.(2) or its DFT equivalent merely evaluates the static Coulomb interaction between the electron and hole, and neglects the kinetic energy of the hole, thus yielding only an approximate $E_A$. If the kinetic energy of the hole or the $k \neq 0$ component of the Coulomb potential is taken into account, as in Eq.(1), the Coulomb contribution will not be as simple as that only between $\varphi_I$ and $\varphi_{VBM}$, which will be discussed later. The most important message of Eq.(2) is that $\delta E_{tot}$ and $E_I$ are two different physical quantities.

Similarly we can write the free-to-bound transition energy as the total energy difference between the two states of the system:

$$E_{F-B} = E_{CBM} - E_I - \left[\langle \varphi_I \varphi_{CBM} \left|\frac{e^2}{\varepsilon r}\right| \varphi_I \varphi_{CBM} \rangle - \langle \varphi_I \varphi_{CBM} \left|\frac{e^2}{r}\right| \varphi_{CBM} \varphi_I \rangle \right]. \qquad (3)$$

Therefore, $E_{F-B} = E_g - E_I - E_{A'}$, where $E_{A'}$ is given by the terms in the square brackets for the Coulomb contribution involving the CBM instead of the VBM in $E_A$. $E_{A'}$ is expected to be in the order of free exciton binding energy, and will not be so significant if $E_g - E_I$ is relatively large. Therefore, $E_g - E_I$ could be taken as an approximation for $E_{F-B}$ in the situations where $E_g - E_I \gg E_{A'}$ is valid.

We will discuss below the three representative approaches that can be found in the literature for computing the transition energies associated with the acceptor within the framework of DFT. Rather than trying to judge which method is more accurate, our intent here is to highlight the different meanings of the results obtained from these different approaches. We will use Si:In as a prototype system to illustrate the differences [4]. The comparison is made for the results all obtained within the local density approximation (LDA). Despite the limitation imposed by the LDA in the accuracy of the absolute transition energies, these results are sufficient to serve the purpose – revealing the differences in the underlying physics.

**(1) Total energy difference between the excited and ground state**

In the literature, the total energy difference $\delta E_{tot}$ is commonly used or implied as the quantity to be compared with the experimentally derived "acceptor binding energy" $E_A^{exp}$ or as a more accurate version of the acceptor binding energy $E_A$ in Eq.(1). However, as pointed out above, $\delta E_{tot}$ is actually an approximate value for the transition or formation energy of the acceptor bound exciton, $E_{A,ex}$, thus should not be compared with $E_A^{exp}$ that actually measures the single particle energy $E_I$, and neither should it be viewed as $E_A$ that describes the Coulomb interaction.

There are actually two different ways to calculate $\delta E_{tot}$. The conceptually most straightforward way to evaluate the transition energy of the whole system between the excited and ground state should be, with the total number of the valence electrons (N) fixed, calculating the total energy difference between them with one electron being removed from the VBM and forced to occupy the $E_I$ level in the excited state (the so-called constrained DFT or selective occupation). The result is referred to as $\delta E_{tot}^N$. However, more often in the literature, the excited state of the system is simulated by a system with one extra valence electron added to the original system or (N+1) valence electrons, where simultaneously a uniform positive background is introduced to compensate the charge of the extra electron [15, 16]. The result may be referred to as $\delta E_{tot}^{N+1}$. There are some subtle differences between the two methods. Some brief comments will be offered at the end of this section. Nevertheless, either way, this total energy difference approach yields an approximation for $E_I - E_A$. Because the kinetic energy of the hole is neglected, $E_A$ is potentially over estimated, resulting in a smaller transition energy $E_{A,ex}$. To correctly describe the Coulomb contribution and explain those abundant discrete transitions in absorption [29], one would need to convert Eq.(2) into an excitonic equation (also known as a Bethe-Salpeter equation) by taking into account the kinetic energy of the hole [34, 36]. If this last step is carried out, we should have the most rigorous treatment for the acceptor problem. A simplified treatment of the excitonic problem will be given later along with the second DFT based approach.

Taking Si:In as an example, the DFT-LDA calculations yielded $\delta E_{tot}^{N+1}$ = 39 meV [35], $\delta E_{tot}^N$ = 36 meV [4]. Apparently, δE$_{tot}$ is much smaller than $E_A^{exp}$ = 153 meV. Besides the limitation of the computational method, for instance, the LDA, which tends to result in a smaller transition energy, one should note that $\delta E_{tot}$ and $E_A^{exp}$ represent two different physical quantities that should not be directly compared with each other. The distinction will be clearer after the discussions to be given for the other two DFT based approaches.

**(2) Total energy calculation of the ground state**

By performing only the ground state calculation (with N electrons), one can obtain the neutral impurity state E$_I$ and its wave function φ$_I$. One can go one step further to solve the whole bound exciton problem. This problem is similar to the well-known free exciton problem where the excitonic states can be further calculated after the one-electron band structure is obtained with the system in the ground state. If the Coulomb interaction is relatively weak, the Coulomb contribution can be described by an effective mass equation with the point charge in Eq.(1) replaced by a charge density d(r) and an exchange term, as given below for an isotropic and parabolic single valence band [28]:

$$\left(\frac{\hbar^2}{2m_h^*}\nabla^2 + \frac{d(r)e^2}{\epsilon r} - J\rho(r)\right) F(r) = E_A^{eff} F(r), \tag{4}$$

with d(r)/r = $\Sigma_\mathbf{k}$exp($i\mathbf{k}\cdot\mathbf{r}$)f(k)s($\mathbf{k}$) being the Coulomb potential with its Fourier component f(k) weighted by s($\mathbf{k}$) = $\Sigma_{\mathbf{k}'}a^*(\mathbf{k}')a(\mathbf{k}'-\mathbf{k})$, where $a(\mathbf{k})$ is the $\mathbf{k}$ component of the impurity wave function φ$_I$ expanded in the basis of the bulk states; J is approximately the exchange term in Eq.(2), and ρ(r) = $\Sigma_\mathbf{k}$exp($i\mathbf{k}\cdot\mathbf{r}$)s($\mathbf{k}$) ≈ |φ$_I$|$^2$. Apparently, if |φ$_I$|$^2$ is a δ function, we have d(r) = 1, and Eq.(4) is essentially the same as Eq.(1). Because of the finite extension of the impurity state, the binding energy will be smaller than $E_A^{eff}$ from the idealistic effective mass equation, which is exactly what has been observed experimentally for the dependence of the hole binding energy of the bound exciton on the electron binding energy: $E_h$ depends on the electron binding energy but always $E_h < E_A^{eff}$ for all NN$_i$ and N centers in GaP [26, 28]. The reduction was initially interpreted as due to some unspecified central cell correction [26], but now is more correctly explained as due to the finite extension of the electron bound state [28]. As an approximation, one could neglect the finite extension of the $E_I$ state or skip Eq.(4) by simply taking the multi-band effective mass solution as an upper bound of $E_A$ [12]. Therefore, the burden of solving the acceptor problem lies mostly on the ability of getting the accurate one electron impurity state $E_I$. For Si:In, the calculation based on DFT-LDA has yielded $E_{I,g}^N$ = 49 meV, where "g" stands for "ground state". If taking $E_A \approx E_A^{eff}$ = 27 meV (calculated with the LDA band structure[35]), we have an estimate for the excitonic transition energy for In in Si as $E_{A,ex} \approx E_{I,g}^N - E_A \approx$ 23 meV. $E_{A,ex}$ and $\delta E_{tot}^N$ (= 36 meV) can be viewed as two different approximations for the excitonic transition energy, and are physically different from $E_I$.

This two-step approach is expected to be a reasonably good approximation for solving the acceptor bound exciton problem for many real systems. One potential shortcoming of this approach lies in that it does not account for the difference in the lattice configurations between the excited and ground state. This effect will be examined below.

**(3) Total energy calculation of the excited state**

One may also perform the total energy calculation for an excited state, in particular with the single-electron impurity level being occupied at $E_{I,e}$, where "e" stands for "excited state". This approach basically requires doing the same calculation as in the first approach, with either the N+1 or N electron system, but uses the single particle state to determine the transition energy. For the N+1 system, the DFT-LDA calculation yields $E_{I,e}^{N+1}$ = 58 meV for Si:In [4]. Using this value, the excitonic transition energy is given as $E_{A,ex} = E_{I,e}^{N+1} - E_A^{eff} = 58 – 27 = 31$ meV, which is close to the total energy difference $\delta E_{tot}^{N+1} = 39$ meV. The calculated $E_{I,e}^{N+1}$ value is much smaller than $E_A^{exp} = 157$ meV [13], but the agreement with experiment can be greatly improved after applying GW and other corrections, which yields $E_{I,e}^{N+1} = 139$ meV [37].

In the constrained DFT excited state calculation of the N electron system, $E_{I,e}^{N} = 48$ meV was obtained in LDA, compared to $E_{I,g}^{N} = 49$ meV from the ground state calculation. The difference between $E_{I,e}^{N}$ and $E_{I,g}^{N}$ should be mostly due to the difference in lattice relaxation, which is apparently rather small for Si, but could be larger for other systems. It is worth noting that $E_{I,e}^{N+1} = 58$ meV is noticeably greater than $E_{I,e}^{N} = 48$ meV.

With using either the N+1 or N electron system, after obtaining the impurity state $E_{I,e}$, in order to account for those discrete absorption features observed experimentally [29], one has to go one step further to treat the excitonic problem as in Eq.(4) or in a more rigorous manner beyond the effective mass approximation.

There is clearly a qualitative correlation between the $E_I$ energy calculated by DFT and the p-orbital energy of the valence electron with respect to the Si 3p orbital for Si:III [19, 35, 37], as shown in Fig. 6, which is consistent with our understanding about the nature of the impurity state. As a matter of fact, the spatial extension of the impurity state wave function, plotted by spherically averaged radial distribution of $|\varphi_I|^2$, is found to be highly localized, and does not resemble at all a hydrogenic state, for all the group III elements, including the shallowest acceptor B [35]. One might be tempted to interpret this wave function localization in terms of the "central cell correction" to Eq.(1). However, we should realize that $E_I$ fundamentally is an eigenvalue of the single particle Kohn-Sham equation that will never produce the abundant discrete absorption lines in the IR absorption spectrum of an acceptor, as shown in Fig. 9 for Si:Ga and similarly in Fig. 8 for GaP:N, because it does not address the excitonic nature of the acceptor problem.

Table 1 summarizes the results of the three different approaches. The numerical results are qualitatively and more or less quantitatively consistent, considering the variations in computational details and approximations involved.

**Table 1.** DFT-LDA results for Si:In (in meV). "e" – excited state, "g" – ground state of the system. The first lines are the results of Ref.[4], the second lines of Ref.[35].

| Approach<br>Total Energy Calc. | (1)<br>Total energy difference | (2)<br>Single particle state of system ground state | (3)<br>Single particle state of system excited state |
|---|---|---|---|
| Constrained DFT<br>$E_{tot,e}(N)$, $E_{tot,g}(N)$ | $E_{A,ex} \approx \delta E_{tot}^N = 36$ | $E_{I,g}^N = 49$<br>$E_A^{eff} = 27$<br>$E_{A,ex} = 22$ | $E_{I,e}^N = 48$<br>$E_A^{eff} = 27$<br>$E_{A,ex} = 21$ |
| $E_{tot}(N+1)$, $E_{tot}(N)$ | $E_{A,ex} \approx \delta E_{tot}^{N+1} = 39$ | (should be the same as above) | $E_{I,e}^{N+1} = 58$<br>$E_A^{eff} = 27$<br>$E_{A,ex} = 31$ |

Finally, some brief comments are provided about the subtle differences between the DFT calculations with conserved or non-conserved total electron numbers. It is customary in the literature to change the total number of valence electrons in the system to emulate different charge states. For instance, in the DFT calculation for the transition from a neutral vacancy state $V^0$ to an ionized vacancy state $V^+$, the $V^+$ state is simulated by a system with one less valence electron, plus a uniform negative background charge equivalent to one electron. One may understand the uniform background charge representing the plane wave state of the electron or corresponding to the case where the electron has been excited to the vacuum level, which is perhaps more relevant to the photoemission experiment. An alternative more relevant to the interband transition in a semiconductor is to let the electron occupy the CBM, mimicking the photoexcitation from the defect to conduction band transition [3]. In the single particle picture, the energy levels do not depend on which states are involved in the transition. Thus, the two options should not make any major difference. However, in the many-electron self-consistent calculation, the atomic configuration of the impurity does depend on the charge distribution of the neighboring atoms, which is why the two options could potentially make some practical difference for the case of strong lattice relaxation. For the case of $A^0$ to $A^-$ transition for an acceptor, taking one electron from the VBM to the impurity state at $E_I$ seems to be most relevant to either IR absorption or photo-conductivity measurement for the acceptor. The more commonly adopted approach, adding one extra valence electron plus a uniform positive background charge, could be problematic, because on one hand the situation conceptually resembles an electron affinity calculation (if we do not consider the added background charge), which tends to yield a larger transition energy; on the other hand, if the background charge were viewed as a hole state, clearly it would not be a good approximation for any real valence band state. The results of Table 1 allow us to examine the real effects. The fact that $E_{I,e}^{N+1}$= 58 meV is greater than $E_{I,e}^{N}$= 48 meV reflects the difference between the two options. Also, while the impurity binding energy does not change much between the ground state and excited state, $E_{I,g}^{N}$= 49 meV vs. $E_{I,e}^{N}$= 48 meV using the N electron selective occupation scheme, the difference is significantly larger between the N and N+1 electron scheme: $E_{I,e}^{N}$= 48 meV vs. $E_{I,e}^{N+1}$= 58 meV. The contrast between the two schemes can be understood in terms of the change in charge distribution near the defect site. With the selective excitation the charge distribution and thus the lattice relaxation is expected to be smaller, because an electron is moved from the p-like VBM to the similar p-like impurity state, whereas with the N + 1 scheme, the charge distribution change is likely more drastically because now both p-like VBM and impurity state are occupied, thus, the repulsion of the two p-like states tends to yield a higher impurity state. Therefore, the selective excitation scheme is preferred for problems related to the electronic transitions involving the impurity and bulk states.

## 7. Summary

The conventional hydrogen model for "shallow" impurities overlooks the impurity state that is typically a highly localized state and instead only focuses on the Coulomb interaction between the ionized impurity core and the excited carrier. The consequence of the Coulomb interaction is interpreted mistakenly as the donor or acceptor binding energy. In the new model, the distinction between the "deep" and "shallow" impurities essentially disappears. They all can be understood under a unified framework of the bound exciton model, although with some subtle differences. This new understanding implies that many existing experimental data in the literature should be re-analyzed and explained, and it can also have real impact on device design.

The results of different first-principles based impurity calculations may mean different things, depending on which approach is adopted. In the total energy approach, the total energy difference between the excited and ground states gives approximately the transition energy of the bound exciton state, which is smaller than the activation energy of the free carrier electrical conductivity. The single particle state instead should in principle yield the impurity state or the impurity binding energy that is directly relevant to the free carrier electrical conductivity, which, however, is conceptually irrelevant to what is described by the hydrogen mode of the conventional theory. Furthermore, there is subtle but important differences between using selective occupation and uniform background charge in calculating the defect states.

**Acknowledgment**

This work was supported by ARO/MURI (W911NF-10-1-0524 monitored by Dr. William Clark), ARO/Electronics (W911NF-16-1-0263 monitored by Dr. William Clark and Dr. Michele